\documentclass[
preprintnumbers,
amsmath,amssymb]{revtex4}
\usepackage[colorlinks=true, pdfstartview=FitV, linkcolor=blue, citecolor=red, urlcolor=magenta]{hyperref}
\usepackage{amssymb}
\usepackage{amsmath}
\usepackage{graphicx}
\usepackage{hyperref}
\usepackage{latexsym}
\usepackage{epsfig}
\usepackage{color}
\allowdisplaybreaks
\begin{document}
\title{Exploring the Effects of Generalized Entropy onto Bardeen Black Hole Surrounded by Cloud of Strings}
\author{Hamza Tariq $^{1}$ \footnote{raohamzatariq4@gmail.com}}
\author{Usman Zafar $^{2}$ \footnote{zafarusman494@gmail.com; usmanzafar366@yahoo.com}}
\author{Shahid Chaudhary$^{3, 4}$
\footnote{shahidpeak00735@gmail.com}}
\author{Kazuharu Bamba $^{2}$\footnote{bamba@sss.fukushima-u.ac.jp}}
\author{Abdul Jawad $^{5,6}$
\footnote{jawadab181@yahoo.com; abduljawad@cuilahore.edu.pk}}
\author{Sanjar Shaymatov $^{7,8,9}$\footnote{sanjar@astrin.uz}}
\address{$^1$ Department of Mathematical Sciences, Faculty of Science, Universiti Teknologi Malaysia, 81310 Johor Bahru, Johor, Malaysia}
\address{$^2$~Faculty of Symbiotic Systems Science, Fukushima University,
Fukushima 960-1296, Japan.}
\address{$^3$ Department of Natural Sciences
and Humanities, University of Engineering and Technology Lahore, New
Campus, Pakistan}
\address{$^4$Research Center of Astrophysics and
Cosmology, Khazar University, Baku, AZ1096, 41 Mehseti Street,
Azerbaijan}
\address{$^5$ Department of Mathematics, COMSATS University Islamabad,
Lahore-Campus, Lahore-54000, Pakistan.}
\address{$^6$~Institute for Theoretical Physics and Cosmology,\\ Zhejiang
University of Technology, Hangzhou 310023, China}
\affiliation{$^7$ University of Tashkent for
Applied Sciences, Str. Gavhar 1, Tashkent 100149, Uzbekistan.}
\affiliation{$^8$ Shahrisabz State Pedagogical Institute, Shahrisabz Str.
10, Shahrisabz 181301, Uzbekistan.} \affiliation{$^9$ Western Caspian
University, Baku AZ1001, Azerbaijan.}

\date{\today}

\begin{abstract}
This work explores the thermodynamic characteristics and geothermodynamics
of a Bardeen black hole (BH) that interacts with a string cloud and
is minimally connected to nonlinear electrodynamics. To avoid the
singularities throughout the cosmic evolution, we consider an
entropy function which comprises five parameters. In addition, by
employing this entropy function for the specific range of
parameters, we obtain the representations of BH entropy based on the holographic principle. Moreover, we employ this entropy function to
investigate its impact on the thermodynamics of the BH by studying
various thermodynamic properties like mass, temperature, heat
capacity, and Gibbs free energy for numerous scalar charge and string cloud values. To support our investigation, we use various
geothermodynamics formalisms to evaluate the stable behavior and
identify different physical scenarios. Furthermore, in this
analysis, we observe that only one entropy formalism provides
us with better results regarding the thermodynamic behavior of the BH. Moreover, it is shown that one of the entropy models provides a thermodynamic geometric behavior compared to the other entropy models.  \\


\end{abstract}

\maketitle

\section{Introduction}

The exploration of the black hole (BH) thermodynamics
\cite{Gibbons:1977mu, Gibbons:1996af} inspired by Hawking's
identification of thermal radiation emitted by BH
\cite{Hawking:1975vcx} is highly relevant for various reasons. For
example, thermodynamics becomes crucial for comprehending the
complex, large-scale operations of the universe in the field of
cosmology, where the study includes contributions from various
galaxies and stars. Another reason is that if there is no entropy in
classical BH, it challenges the second law of BH thermodynamics.
Moreover, to fully clarify the behavior of the BH, the inclusion of quantum
effects becomes necessary to underscore the present deficiencies in our comprehension of quantum gravity and the merging of total
fundamental forces. The revelation that BH radiation possesses a finite temperature and aligns with the Hawking-Bekenstein entropy function \cite{Hawking:1975vcx, bekenstein1973black} is viewed as a substantial achievement in theoretical physics function. The
Hawking-Bekenstein entropy is notably proportional to the horizon
area of a BH in contrast to classical thermodynamics, where entropy
typically scales with the volume of the system. This unique
characteristic has led to the development of alternative entropy
functions such as Tsallis \cite{tsallis1988possible} and R\'{e}nyi
entropies \cite{odintsov2023non}, which account for the system's non-additive statistics. Recently,~\cite{Barrow:2020tzx}
introduced an entropy function that incorporates the fractal structure of BH, which is potentially linked to quantum gravity effects. Several additional entropy models, such as Sharma-Mittal (SM) entropy, Kaniadakis entropy \cite{das2021quantum,calcagni2017stability}, and Loop Quantum Gravity entropy \cite{rovelli1996black, barbero2024black}, each detailed in respective Refs.~\cite{jahromi2018generalized,kaniadakis2005statistical,majhi2017non, liu2022non}, share the characteristic of simplifying into the Hawking-Bekenstein entropy under certain conditions and are consistently increasing functions relative to the Hawking-Bekenstein entropy variable.
A significant amount of work has been conducted to examine the thermodynamics of black holes and analyze their critical behavior and phase transitions, as discussed in Refs.~\cite{Mandal:2016anc,Chamblin:1999tk, Chamblin:1999hg, Hendi:2012um, Yazdikarimi:2019jux, Guo:2019oad, Appels:2016uha, Appels:2017xoe, Gunasekaran:2012dq, Deng:2018wrd}.

In this context, the concept of Tsallis, R\'{e}nyi, and SM entropies
becomes notable. Serving as an alternative measure of entropy, it
provides valuable perspectives on the thermodynamic characteristics
of intricate systems, including BH. The application of Tsallis,
R\'{e}nyi and SM entropy in BH thermodynamics opens up an engaging
pathway to investigate these celestial objects' statistical properties and information
content. In
Ref.~\cite{jahromi2018generalized} proposed that the SM entropy
formed by combining the Tsallis and R\'{e}nyi entropies yields
fascinating outcomes within the cosmological framework. Furthermore,
geometrical thermodynamics serves as a robust framework for
examining the phase transition of BH, leading to various
thermodynamic metrics. By formulating the thermodynamic metric
regarding the entropy, its divergence points of curvature scalar
offer crucial insights into potential phase transitions within the
BH system. Initially, Weinhold ~\cite{weinhold1975metric} proposed the
metric formalism based on the equilibrium state space of the
thermodynamic systems. However, Ruppeiner
\cite{Ruppeiner:1979bcp, ruppeiner1995riemannian} created
different metric formalisms that showed an equal and compatible
association with the Weinhold metric. Moreover, it has been noted
that these aforementioned metrics are not invariant when subjected
to a Legendre transformation. Recognizing these limitations, Quevedo
\cite{Quevedo:2006xk, rani2024thermodynamic} introduced the
first metric with Legendre invariance to address the issues
associated with the preceding two metrics. However,
the Quevedo metric only partially proves to be a successful model in numerous particular systems as its Ricci scalar
exhibits additional divergence points without clear physical interpretation. Ultimately,
Refs.~\cite{Hendi:2015rja, Hendi:2015xya} offers a metric
formalism, where the problem of mismatched divergence is not evident
\cite{Soroushfar:2020wch, chabab2019phase}. Numerous works have been done by employing various thermodynamic geometry formalisms (for further details check Refs.~\cite{Ruppeiner:2008kd, Sahay:2010tx, Lala:2011np, Li:2016wzx, Sahay:2016kex, KordZangeneh:2017lgs, Soroushfar:2019ihn, Bhattacharya:2019qxe, Kumara:2019xgt}).

 Fundamentally, photons released from a bright source
near a BH face two potential fates, either yielding to the
irresistible gravitational attraction and steadily approaching the
event horizon or being redirected away, embarking on an endless
journey into the expansive cosmic realm. The intricate cosmic dance
delineates crucial geodesic paths, identified as unstable spherical
orbits, which mark the boundary between these two possibilities.
Through precise observation of these critical photon trajectories
against the cosmic backdrop, we gain the remarkable ability to
capture the unseen visuals of a BH shadow \cite{Filho:2023abd}. In BH thermodynamics, researchers turn to Tsallis, Rényi, and Sharma-Mittal (SM) entropies as a way to go beyond the usual Boltzmann-Gibbs entropy. These alternatives help capture the complexities of systems where interactions don’t behave in the standard way—where correlations, non-extensive effects, and even quantum gravity start to play a critical role (for more details regarding the idea of non-extensive entropies and their application in cosmology and BH thermodynamics see Refs.~\cite{Nojiri:2022aof,Nojiri:2022dkr,Nojiri:2022sfd,Elizalde:2025iku,Odintsov:2023vpj,Nojiri:2024zdu}). Tsallis entropy naturally fits into non-extensive statistical mechanics, making it a powerful way to understand systems where long-range interactions, like gravity, play a major role \cite{tsallis1988possible}. Rényi entropy offers a unique way to describe systems that don't follow traditional additive rules, tweaking additivity in a controlled manner \cite{odintsov2023non}. This makes it especially useful in quantum information theory when studying BH entropy. Moreover, SM entropy offers a more flexible and comprehensive approach than Tsallis and Rényi entropies, allowing for extra tuning parameters that help account for quantum gravitational effects and deviations from conventional thermodynamics \cite{jahromi2018generalized, Drepanou:2021jiv}. These ways of measuring entropy are especially important when studying BHs in settings that include quantum effects, non-traditional behaviors, and holography. They help us gain a deeper understanding of what’s really happening at a microscopic level, going beyond the usual Bekenstein-Hawking perspective.

Thereby, the above-mentioned are the major reasons that motivated us to choose Tsallis, R\'{e}nyi, and SM entropies because these entropies are good at describing the unique properties of BH thermodynamics. Tsallis entropy helps us understand long-range interactions, Rényi entropy offers flexibility in describing different states, and SM entropy combines both approaches to comprehensively investigate the thermodynamics of BHs. Our findings show new connections between specific heat capacity and Gibbs free energy, indicating possible phase changes in BH, which improves our understanding of their thermodynamic behavior. This research paper aims to examine the thermodynamic
properties of BH using the Tsallis, R\'{e}nyi, and SM entropy
framework. We investigate the consequences by applying these entropies measures to the above-mentioned BH solution, with a specific emphasis
on its ability to provide insights into the statistical behavior and information processing capacities of BH.

The structure of the paper is outlined as follows: Section II
presents a summary of the approach used to integrate Tsallis,
R\'{e}nyi and SM entropy into BH thermodynamics. In Section III, we
explore the thermodynamic characteristics of these BH, investigating
parameters such as temperature, specific heat, the Gibbs free energy,
and other thermodynamic values. Section IV discusses the
thermodynamical geometries of Tsallis,  R\'{e}nyi, and SM entropies.
In Section V, we have some concluding remarks and discuss possible
directions for further investigation.

\section{Thermodynamics of a Bardeen Black Hole Surrounded by a String Cloud}

We consider the action associated with General Relativity (GR), which
exhibits minimal coupling to non-extensive dynamics (NED) involving the action about a string cloud is given by \cite{Rodrigues:2022zph}
\begin{equation}\label{401}
I=\int \bigg\{\mathcal{L}(F)+ 2\lambda+R\bigg\} \sqrt{-g}\ d^{4} x
+I_{CS},
\end{equation}
where $g,~R,~\lambda$ represents the metric determinant, Ricci
curvature scalar and cosmological constant, respectively.
Furthermore, $\mathcal{L}(F)$ signifies the non-linear Lagrangian
describing electromagnetic theory, and it is dependent on the scalar
function $F=\frac{F^{\alpha\beta}F_{\alpha\beta}}{4}$,
where
$F_{\alpha\beta}=-\partial_{\beta}A_{\alpha}+\partial_{\alpha}A_{\beta}$
is defined as electromagnetic field intensity $A_{\alpha}$ \cite{hyun2019charged}.
Moreover, the last term in our action is $I_{CS}$, which is
Nambu-Goto action and it is employed for describing string-like
entities and it is expressed as follows,
\begin{equation}\label{402}
I_{CS}=\int  \sqrt{-\gamma} M d \Lambda^{0}d \Lambda^{1},
\end{equation}
where $\gamma$ in Eq.~(\ref{402}) denotes the determinant of the
induced metric $\gamma_{AB}$ on the submanifold, as explicitly
stated by
\begin{equation}\label{403}
 \gamma_{AB}=g_{\alpha\beta}\frac{\partial x^{\alpha} \partial x^{\beta}}{\partial \Lambda^{A} \partial
 \Lambda^{B}}.
\end{equation}
Here, $\Lambda^{0}$ and $\Lambda^{1}$ represent parameters
characterizing the time-like and space-like attributes of the
system, respectively while $M$ represents a constant with no
dimension, which describes the string. Thereby, the action Eq.
(\ref{401}) is subjected to variation with respect to the metric
tensor $g_{\alpha\beta}$ to obtain
\begin{equation}\label{404}
R_{\alpha\beta}+g_{\alpha\beta}
\lambda-\frac{1}{2}g_{\alpha\beta}R=8 \pi T_{\alpha \beta}+8 \pi
T^{cs}_{\alpha\beta},
\end{equation}
where $T_{\alpha\beta}$ denotes the energy-momentum tensor (EMT)
associated with the matter sector in the case of NED and
$T^{cs}_{\alpha\beta}$ represents the EMT specific to the string
cloud tensor. One can easily define these tensors as follows
\begin{equation}\label{405}
 T_{\alpha\beta}=g_{\alpha\beta} L(F)-\frac{ d L}{d F}
 F^{\gamma}_{\alpha} F^{\beta}_{\gamma},
 T^{cs}_{\alpha\beta}=\frac{\sigma \sum^{\gamma}_{\alpha} \sum_{\gamma\beta}}{8 \pi
 \sqrt{-\gamma}},
\end{equation}
where $\sigma$ in the above equation is the cloud proper density
while $\Sigma^{\alpha\beta}$ is a bi-vector. Thus, the spherically
symmetric spacetime, in this case, is given as
\begin{equation}\label{406}
  ds^{2}=f(r)dt^{2}-\frac{1}{f(r)}d r^{2}-r^{2}d
  \theta^{2}-r^{2}\sin^{2}\theta d\phi^{2},
\end{equation}
where the metric function is denoted by $f(r)$ while for the Bardeen
solution its Lagrangian is given by
\begin{equation}\label{407}
  \mathcal{L}(F)=\frac{3}{8 \pi s q^{2}}\bigg(\frac{\sqrt{2 q^{2}F}}{2+\sqrt{2
  q^{2}F}}\bigg)^{5/2}.
\end{equation}
Here, $S=\frac{\mid q \mid}{2M}$,  $q$ represents the magnetic
monopole charge while $M$ is the mass of the BH.
Upon resolving the field equation, one can compute the set of
distinct differential equations, which are non-trivial and it is given as
\begin{equation}\label{408}
   \frac{\lambda r^{2}-f(r)-r f^{'}r+1}{r^{2}}-\frac{6
   M}{q^{3}(\frac{r^{2}}{q^{2}}+1)^{5/2}},
\end{equation}
\begin{equation}\label{409}
  \lambda- \frac{f^{'}(r)}{r}-\frac{f^{''}(r)}{2}- \frac{ \sqrt{\frac{q^{2}}{r^{2}}+1}\Big(-5 q^{10}+2 q^{2}r^{8}+2
  r^{10}\Big)3 M}{q^{3}r^{2}(q^{2}+r^{2})^{4}}=0.
\end{equation}
The prime symbol represents differentiation with respect to $r$,
while $a$ represents an integration constant related to strings,
constrained by the range of $a$ which is (0,~1). Further details
regarding the preceding discussion can be found in Ref.~\cite{Rodrigues:2022zph} (It is also shown that the Bardeen BH can arise as a particular case of shift and parity symmetric Horndeski theories in Ref.~\cite{Bakopoulos:2024ogt}). Therefore, one can compute the metric
function by solving the Eqs.~(\ref{408}) and (\ref{409}) which
yields
\begin{equation}\label{410}
   f(r)=1-\lambda \frac {r^{2}}{3}-a-\frac{2 M
   r^{2}}{(q^{2}+r^{2})^{3/2}}-\frac{2 \omega}{r},
\end{equation}
where $\omega$ is an integration constant, and when we set
$\omega=0$, the solution mentioned above transforms into the
solution of Bardeen-AdS.
\begin{equation}\label{411}
   f(r)=1-\lambda \frac {r^{2}}{3}-a-\frac{2 M
   r^{2}}{(q^{2}+r^{2})^{3/2}}.
\end{equation}

Moreover, excluding the string parameter $a$ from the
aforementioned solution results in the original Bardeen
solution. Entropy is a fundamental concept in physics that varies with the specific characteristics of the physical system under
investigation. For instance, in classical thermodynamics, entropy is
linked to the system volume, while for the BH, it is proportional to
the area of the event horizon. This suggests that our current
comprehension of entropy's fundamental nature may be insufficient, or
perhaps a more comprehensive form of entropy exists that
applies universally across diverse systems. The discovery of BH
radiation in theoretical physics is highly significant, which is characterized by a certain temperature and is dictated by the
Hawking-Bekenstein entropy function \cite{
Hawking:1975vcx, Bekenstein:1980jp} (for more details, see Refs.~\cite{Bardeen:1973gs,
Wald:1999vt}). What sets the Hawking-Bekenstein entropy apart is its
direct proportionality to the area of the BH event horizon, unlike
classical thermodynamics, where entropy scales with the volume of the
system. This unique feature of BH entropy has induced the
development of various alternative entropy functions. Examples
include the Tsallis entropy \cite{tsallis1988possible}  and the R\'{e}nyi entropy \cite{odintsov2023non}, which incorporate non-additive statistics of the system. Recently, Ref.~\cite{Barrow:2020tzx}
suggested an entropy function that accounts for the fractal structure of BH, which is possibly influenced by quantum gravity effects. Other significant forms of entropy are briefly discussed in Refs.~\cite{jahromi2018generalized, Drepanou:2021jiv}.
These entropy definitions all converge to the Hawking-Bekenstein
entropy under specific conditions and exhibits a monotonic increasing
behavior relative to the Hawking-Bekenstein entropy variable. We
consider a novel entropy function that is free from singularities
\cite{odintsov2023non}, and defined as
\begin{eqnarray}\label{412}
S [\alpha_{\pm} ,\beta ,\gamma ,\epsilon] =
\frac{1}{\gamma}\Bigg[\Bigg\{{1+\frac{1}{\epsilon}
\tanh\Bigg(\frac{\epsilon
\alpha_{+}}{\beta}S_{o}\Bigg)}\Bigg\}^{\beta} -
\Bigg\{1+\frac{1}{\epsilon}   \tanh\  \Bigg(\frac{\epsilon
\alpha_{-}}{\beta}
  S_{o}  \Bigg)  \Bigg\}^{-\beta}\Bigg],
\end{eqnarray}
where \( S_{0} = \frac{1}{2}\pi^{2}r_{+}^{3} \) denotes the
Hawking-Bekenstein entropy and \( A_{\pm}, B, \gamma \), and \(
\epsilon \) are positive constants. Utilizing the value of \( S_{0}
\), the generalized entropy is obtained as
\begin{eqnarray}\label{413}
S [\alpha_{\pm} ,\beta ,\gamma
,\epsilon]&=&\frac{1}{\gamma}\Bigg(\Bigg[{1+\frac{1}{\epsilon}
\tanh\Bigg\{\frac{\epsilon \alpha_{+}}{\beta}\Bigg(\frac{1}{2}
\pi^{2}r^{3} \Bigg) \Bigg\}}\Bigg]^{\beta}-
 \Bigg[1+\frac{1}{\epsilon}
\tanh   \Bigg\{ \frac{\epsilon \alpha_{-}}{\beta}\Bigg(\frac{1}{2}
\pi^{2} r^{3}\Bigg)\Bigg\}
\Bigg]^{-\beta}\Bigg).
\end{eqnarray}
As \(\epsilon \rightarrow 0\), \({\alpha_{+} \rightarrow \infty}\),
\({\alpha_{-} \rightarrow 0}\), and \(\gamma =
\left(\frac{\alpha_{+}}{\beta}\right)^\beta \), the generalized
entropy \(S\) reduces to the following form
\begin{equation}\label{414}
S=\Bigg(\frac{1}{2} \pi^{2} r^{3}\Bigg)^{\beta},
\end{equation}
which resembles the Tsallis entropy when $\beta =\delta$
\cite{tsallis1988possible}. By setting \(\epsilon \rightarrow 0\),
\(\beta \rightarrow 0\), \(\alpha_{-} = 0\),
\(\frac{\alpha_{+}}{\beta} \rightarrow\) finite,
\(\frac{\alpha_{+}}{\beta} = \alpha\), and \(\gamma = \alpha_{+}\),
the generalized entropy transforms into
\begin{equation}\label{415}
S= \frac{1}{\alpha}\ln \bigg\{1+\alpha \bigg(\frac{1}{2} \pi^{2}
r^{3}\bigg)\bigg\},
\end{equation}
which corresponds to R\'{e}nyi entropy \cite{odintsov2023non}.
Finally, by applying $\alpha_{-}\rightarrow0$,~$\alpha_{+}=R$,
$\gamma=R$ and $\beta=\frac{R}{\delta}$, the
generalized non-singular entropy transforms into the following form
\begin{equation}\label{416}
S=\frac{1}{R}\Bigg\{1+\delta\bigg(\frac{1}{2} \pi ^2
r^3\bigg)^{R/\delta}-1\Bigg\},
\end{equation}
which resembles the SM entropy \cite{jahromi2018generalized}. Now,
from Eq. (\ref{414}), one can get the radius of Tsallis entropy which
is given by
\begin{eqnarray}\label{417}
r=\bigg(2^{\beta } \pi ^{-2 \beta } S\bigg)^{\frac{1}{3 \beta }}.
\end{eqnarray}
Similarly, from Eq. (\ref{415}) we obtain the radius of R\'{e}nyi
entropy is given by
\begin{eqnarray}\label{418}
r=\bigg\{\frac{2\big(-1+e^{S\alpha}\big)}{\pi^{2}\alpha}\bigg\}^{1/3}.
\end{eqnarray}
Again by using Eq. (\ref{416}) one can compute the radius for SM
entropy, which is given by
\begin{eqnarray}\label{419}
r=\left(\frac{2\bigg[-1+\Big\{R\Big(\frac{1}{R}+S\Big)\Big\}^{\delta/R}\bigg]}{\pi^{2}
\delta}\right)^{1/3}.
\end{eqnarray}
\section{Thermodynamics quantities for Tsallis,  R\'{e}nyi and Sharma-Mittal entropies}

\subsection{Mass of the Black Hole}

Incorporating Tsallis, R\'{e}nyi, and SM entropies into the study of
BH thermodynamics deepens our understanding of these mysterious
cosmic objects while offering fresh insights into their statistical
conduct and information processing mechanisms. Subsequently, we
computed the thermodynamic quantities like the Gibbs free energy,
temperature, heat capacity, and also discuss the stability of the
concerning the BH model. In order to discuss the thermodynamic behavior of Bardeen BH, we first obtain the mass by using $f(r)=0$
and then by making some adjustments, one can obtain the mass in terms
of Tsallis entropy, which is given as
\begin{eqnarray}\label{420}
M_{\mathrm{Tsallis}} &=&-\frac{1}{6} I^{-\frac{2}{3 \beta }} \left(I^{\frac{2}{3 \beta
}}+q^2\right)^{3/2} \left(3 a-8 \pi  I^{\frac{2}{3 \beta }}
P-3\right),
\end{eqnarray}
where $I=2^{\beta } \pi ^{-2 \beta } S$. Similarly, by setting
$f(r)= 0$ from Eq.~(\ref{411}) and by using the value of $r$ from
Eq.~(\ref{418}) the expression of mass $(M)$ in terms of R\'{e}nyi
entropy is given by
\begin{eqnarray}\label{421}
M_{\mathrm{Renyi}}&=&-\frac{\pi ^{4/3} \alpha ^{2/3} \left(\frac{2^{2/3} N^{2/3}}{\pi
^{4/3} \alpha ^{2/3}}+q^2\right)^{3/2} \left(3 a-\frac{2^{2/3} 8 \pi
N^{2/3} P}{\pi ^{4/3} \alpha ^{2/3}}-3\right)}{6\ 2^{2/3} N^{2/3}},
\end{eqnarray}
where $N=e^{\alpha  S}-1$. Again, by repeating the same process
which we have done above, but this time, we did this for SM entropy by
using Eqs.~(\ref{411}) and (\ref{419}), it yields
\begin{eqnarray}\label{422}
M_{\mathrm{SM}}&=&\frac{\pi ^{4/3} \delta ^{2/3}}{6\ 2^{2/3} X^{2/3}}
\left(q^2+\frac{2^{2/3} X^{2/3}}{\pi ^{4/3} \delta
^{2/3}}\right)^{3/2} \left(-3 a+\frac{8\ 2^{2/3} P
X^{2/3}}{\sqrt[3]{\pi } \delta ^{2/3}}+3\right),
\end{eqnarray}
where $X=(R S+1)^{\delta /R}-1$.
\subsection{Temperature of the Black Hole}
Since we know that cosmological constant can interpreted as the
pressure and its expression is $P=\frac{-\lambda}{8 \pi}$. Tsallis
temperature is obtained by substituting Eq.~(\ref{420}) in the given
expression $\frac{\partial M}{\partial S}$, and its final form is
given as
\begin{eqnarray}\label{423}
T_{\mathrm{Tsallis}}&=&\frac{I^{-\frac{2}{3 \beta }} \sqrt{I^{\frac{2}{3
\beta }}+q^2} \Bigg\{I^{\frac{2}{3 \beta }} \left(-a+8 \pi
I^{\frac{2}{3 \beta }} P+1\right)+2 (a-1) q^2\Bigg\}}{6 \beta  S}.
\end{eqnarray}
Similarly, we can derive the expression for temperature in the form
of R\'{e}nyi entropy from Eq.~(\ref{421}) which is given as
\begin{eqnarray}\label{424}
T_{\mathrm{Renyi}}=\frac{\sqrt[3]{\alpha } e^{\alpha  S} \sqrt{\frac{2^{2/3}
N^{2/3}}{\alpha ^{2/3}}+\pi ^{4/3} q^2} \bigg\{\sqrt[3]{\pi } (a-1)
\alpha ^{2/3} \left(N_1-N^{2/3}\right)+8\ 2^{2/3} N^{4/3}
P\bigg\}}{6 \pi  N^{5/3}},
\end{eqnarray}
where $N_1=\sqrt[3]{2} \pi ^{4/3} \alpha ^{2/3} q^2$. Now, we
derived the temperature in terms of SM entropy by employing the
Eq.~(\ref{422}) and it turns out
\begin{eqnarray}\label{425}
T_{\mathrm{SM}}&=&\frac{\sqrt[3]{\pi } \delta ^{2/3} (a-1)\left(q^2
X_2-X^{2/3}\right)+8  P(2 X^2)^{2/3}}{6 \pi X^{5/3}} \Bigg\{\sqrt{\pi ^{4/3} q^2+\left(2
X\delta^{-1}\right)^{2/3}} X_1^{\frac{\delta }{R}-1}\Bigg\} \sqrt[3]{\delta},
\end{eqnarray}
where $X_{1}=(R S+1)$ and $X_{2}=\sqrt[3]{2} \pi ^{4/3} \delta
^{2/3} q^2$.
 
Now, by taking the partial derivative of 
M with respect to P, by using Eq.~(\ref{420}) we obtain the volume V for Tsallis entropy is given below,
\begin{eqnarray}\label{426}
V= \frac{\partial M_{\mathrm{Tsallis}}}{\partial P}  =V_{\mathrm{Tsallis}} = -\frac{4}{3} \pi \left( q^2 +  I \right)^{\frac{2}{3}/\beta} )^{\frac{3}{2}}
\end{eqnarray}
Again, by taking the partial derivative of 
M with respect to P, by using Eq.~(\ref{421}) we obtain the volume
V for  R\'{e}nyi  entropy is given below,
\begin{eqnarray}\label{427}
V= \frac{\partial M_{\mathrm{Renyi}}}{\partial P}=V_{\mathrm{Renyi}} = -\frac{4}{3} \pi \left( q^2 + \left( \frac{2^{2/3}  N }{\pi^{4/3} \alpha^{2/3}} \right)^{3/2} \right)
\end{eqnarray}

Similarly again, by taking the partial derivative of 
M with respect to P, by using Eq.~(\ref{422}) we obtain the volume
V for  SM  entropy is given below,
\begin{eqnarray}\label{428}
V= \frac{\partial M_{\mathrm{SM}}}{\partial P}=V_{\mathrm{SM}} = -\frac{4}{3} \Bigg(\frac{\pi \left( q^2 + 2^{2/3}   X \right)^{2/3}) ^{3/2} }{ \left( \pi^{4/3} \delta^{2/3} \right)}\Bigg)
\end{eqnarray}

Now we find pressure by using the expression $\frac{\partial M}{\partial S}$, then the pressure of Tsallis, R\'{e}nyi  and SM entropies  is given in Eqs.~(\ref{429}), (\ref{430}),(\ref{431}) respectively,
\begin{eqnarray}\label{429}
P_{\mathrm{Tsallis}}  = \frac{ 3 \times \left( -1 + a + 
\frac{ r^2  I^{\frac{2}{3} / \beta} }
{ q^2 +  I^{\frac{2}{3} / \beta} } 
\right)^{3/2} 
\left( -3 + 3a - 8  \pi  I^{\frac{2}{3} / \beta} 
\right)^{\frac{2}{3} / \beta} 
}
{ 8 \pi r^2 \left( q^2 + r^2 \right)^{3/2} }
\end{eqnarray}

\begin{eqnarray}\label{430}
P_{\mathrm{Renyi}} = \frac{ 3 \pi^{1/3} \left( -1 + a - 
\frac{ \pi^2 \left( q^2 + \frac{2^{2/3} N^{2/3}}{\pi^{4/3} \alpha^{2/3}} \right)^{3/2} 
\left( -3 + 3a - \frac{8 \cdot 2^{2/3} N^{2/3} }{\pi^{1/3} \alpha^{2/3}} \right) }
{ 3 \left( \pi^{4/3} q^2 + \frac{2^{2/3} N^{2/3}}{\alpha^{2/3}} \right)^{3/2} }
\right) }
{ 8 \cdot 2^{2/3} N^{2/3} \alpha^{2/3} }
\end{eqnarray}

\begin{eqnarray}\label{431}
P_{\mathrm{SM}} = \frac{ 3 \pi^{1/3} \left( -1 + a - 
\frac{ \pi^2 \left( q^2 + \frac{2^{2/3} X^{2/3}}{\pi^{4/3} \delta^{2/3}} \right)^{3/2} 
\left( -3 + 3a - \frac{8 \cdot 2^{2/3} X^{2/3} P}{\pi^{1/3} \delta^{2/3}} \right) }
{ 3 \left( \pi^{4/3} q^2 + \frac{2^{2/3} X^{2/3}}{\delta^{2/3}} \right)^{3/2} }
\right) }
{ 8 \cdot 2^{2/3} X^{2/3} \delta^{2/3} }
\end{eqnarray}

\begin{figure} \centering
\epsfig{file=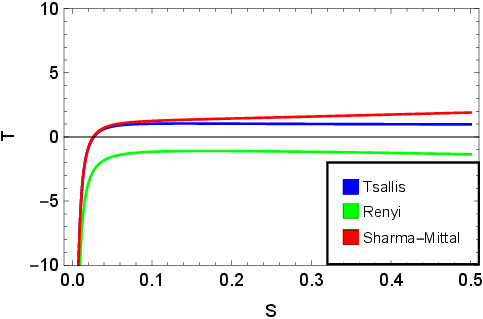,width=.45\linewidth}
\epsfig{file=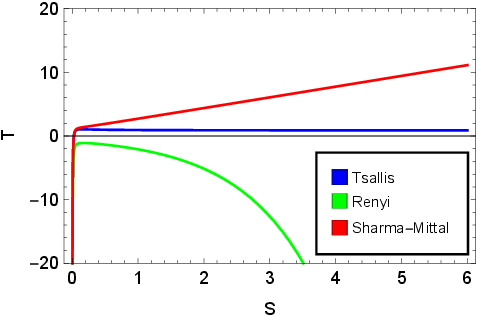,width=.45\linewidth} \caption{$T$
versus Tsallis, R\'{e}nyi is  and  SM entropy $S$ for different values of cloud string $a$ (left panel) and $q$ (right panel) while other parameters are $\beta=1$, $\delta =4$, $R = 2$, $\alpha=0.9$,  and $P = 1$, are held constant. Here, the trajectory of Tsallis, R\'{e}nyi, and  SM entropy are presented in blue, green, and red solid lines, respectively.}\label{Fig-1}
\end{figure}


In this study, we explore the relationships between temperature and
three different generalized entropies like the Tsallis, R\'{e}nyi, and
SM by considering a BH under the influence of a cloud of strings and electric charge. The analysis uses different values of
the cloud of string parameter $a$ and charge $q$, while other
parameters, including $\beta=1$, $\delta =4$, $R = 2$, $\alpha=0.9$,  and $P = 1$, are held constant. Fig.~\ref{Fig-1} (left panel) depicts the relationship between temperature $T$ and Tsallis, R\'{e}nyi, and
SM entropies for different values of $a$. The Tsallis (blue curve) and SM (red curve) entropies show positive behavior, which indicates the physical behavior of BH, while R\'{e}nyi (green curve) shows negative behavior, which demonstrates that R\'{e}nyi entropy shows the non-physical behavior of BH. As the entropy increases, the temperature of SM (red curve) rises, and the graph becomes more positive. It suggests that higher
entropy values correspond to higher temperatures, reinforcing the
physical interpretation of the BH. Temperature initially shows negative behavior for lower entropy values; as the value of entropy increases, the temperature eventually starts increasing. Tsallis and SM become positive, but R\'{e}nyi (green curve) entropy remains negative, indicating that the BH exhibits non-physical behavior. We conclude that  SM (red curve) shows more physical behavior than Tsallis (blue curve) entropy, and R\'{e}nyi (red curve) entropy shows non-physical behavior for different values of $a$. In Fig.~\ref{Fig-1} (right panel), we show a graph that compares Temperature with Tsallis, R\'{e}nyi, and SM entropies for different values of $q$. The Tsallis (blue curve) and SM (red curve) entropies show positive values, indicating that the BH behaves physically across different values of $q$. On the other hand, the R\'{e}nyi (green curve) entropy shows negative values, which suggests the BH behaves non-physically under these conditions. The graph helps us understand that the BH has different thermodynamic behavior depending on the type of entropy used. Tsallis (blue curve) and SM (red curve) entropies show that the BH stays in a physical state as temperature and entropy increase, making them useful for describing BH thermodynamics in situations involving a cloud of strings and electric charge. In contrast, R\'{e}nyi (green curve) entropy behaves differently, showing non-physical behavior, particularly when considering the cloud of string and charge parameters. This
difference may indicate limitations in applying R\'{e}nyi entropy to
BH systems in particular physical contexts, specifically when external fields like strings and charges are considered.

\subsection{Specific Heat Capacity}

In BH thermodynamics, the heat required to change a BH temperature
is called thermal or heat capacity. Heat or
thermal capacity is a key measurable physical property in BH
thermodynamics. The stability of a BH can be determined by its sign,
with a positive sign indicating stability and a negative sign
indicating instability. There are two types of heat capacities: one
that measures the specific heat when heat is added at a constant
pressure $C_{p}$ and another that measures the specific heat when
heat is added at constant volume $C_{v}$. The heat capacity $C_{p}$ \cite{rodrigues2022bardeen, ruppeiner2014thermodynamic,davies1978thermodynamics} is determined by using the  relation
which is given in Eq. (\ref{426})
\begin{eqnarray}\label{432}
C_{p}=\frac{\partial S}{\partial T}\bigg|_{P}T.
\end{eqnarray}
By utilizing the Eqs.~(\ref{414})  and (\ref{423}) into Eq.~(\ref{432}), we can compute the heat capacity $C_{p}$ for the Tsallis entropy case, which is
given by
\begin{eqnarray}\nonumber
C_{p}(\mathrm{Tsallis})&=&\bigg[3 \beta  S \left(I^{\frac{2}{3 \beta}}+q^2\right) \left\{I^{\frac{2}{3 \beta }} \left(-a+8 \pi
I^{\frac{2}{3 \beta }} P+1\right)+2 (a-1)
q^2\right\}\bigg]\bigg[I^{\frac{4}{3 \beta }} \bigg\{a
\\\nonumber&\times&(3 \beta -1)-3 \beta +8 \pi  (2-3 \beta ) P
q^2+1\bigg\}-2 (a-1) (3 \beta +2) q^4-(a-1) I^{\frac{2}{3 \beta
}}\\\label{433}&\times&(3 \beta +2) q^2-24 \pi I^{2/\beta } (\beta
-1) P\bigg]^{-1}.
\end{eqnarray}

Similarly, we obtained $C_{P}$  in the case of R\'{e}nyi entropy by
employing Eqs.~(\ref{415}) and (\ref{424}) in Eq.~(\ref{432}) and it can be written as
\begin{eqnarray}\nonumber
{C_{p}(\mathrm{Renyi})}&=&\bigg[3 N \left(2^{2/3} N^{2/3}+\pi ^{4/3} \alpha
^{2/3} q^2\right) \left\{\sqrt[3]{\pi } (a-1) \alpha ^{2/3}
\left(N_1-N^{2/3}\right)+8\ 2^{2/3} N^{4/3}
P\right\}\bigg]\\\nonumber&\times&\bigg[8 \sqrt[3]{2} \alpha  N P
\left\{2 e^{\alpha S} \left(N_1 \sqrt[3]{N}+3 e^{\alpha S}-6\right)-3
N_1 \sqrt[3]{N}+6\right\}-\sqrt[3]{\pi } (a-1) \alpha ^{5/3}
\bigg\{3\\\nonumber&\times& \left(\pi ^{4/3} \alpha ^{2/3} N^{2/3}
q^2+2^{2/3} \sqrt[3]{N}+\pi ^2 \alpha ^{2/3} q^2\right)+e^{\alpha S}
\big(2 \pi ^{4/3} \alpha ^{2/3} N^{2/3} q^2+2 \pi ^2 \alpha ^{2/3}
N_1 q^2\\\label{434}&-&4\ 2^{2/3} \sqrt[3]{N}\big)+2^{2/3}
\sqrt[3]{N} e^{2 \alpha S}\bigg\}\bigg]^{-1}.
\end{eqnarray}
Furthermore, in case of SM entropy, $C_{p}$ is computed by employing
the Eqs.~(\ref{416}) and (\ref{425}) in Eq. (\ref{432}) which takes the following form
\begin{eqnarray}\nonumber
C_{p}{(\mathrm{SM)}} &=&\bigg[3 \delta ^{2/3} X_1 X \left(\pi ^{4/3}
q^2+\frac{2^{2/3} X^{2/3}}{\delta ^{2/3}}\right) \bigg\{\sqrt[3]{\pi }
(a-1) \delta ^{2/3} \left(X_2-X^{2/3}\right)+8\ 2^{2/3} P
\\\nonumber&\times&X^{4/3}\bigg\}\bigg]\bigg[\sqrt[3]{\pi } (a-1) \delta ^{2/3} \bigg\{3
R X \bigg(2^{2/3} X^{4/3}-\pi ^{4/3} \delta ^{2/3} q^2
\left(X^{2/3}+X_2\right)\bigg)-\pi ^{4/3} \delta ^{5/3}
q^2\\\nonumber&\times& \left(X^{2/3}+X_2\right) \left(2 X_1^{\delta
/R}+3\right)-2^{2/3} \delta  \sqrt[3]{X} \left(-4 X_1^{\delta
/R}+X_1^{\frac{2 \delta }{R}}+3\right)\bigg\}-8 \sqrt[3]{2} P X
\bigg\{\delta\\\label{435}&\times& \left(-X_2\right) \sqrt[3]{X}
\left(2 X_1^{\delta /R}-3\right)+3 R X \left(2 X_1^{\delta /R}+X_2
\sqrt[3]{X}-2\right)-6 \delta X^2\bigg\}\bigg]^{-1}.
\end{eqnarray}

\begin{figure} \centering
\epsfig{file=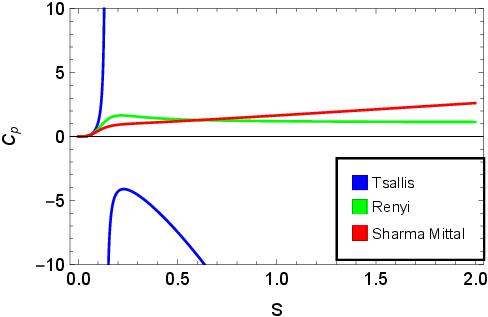,width=.45\linewidth}
\epsfig{file=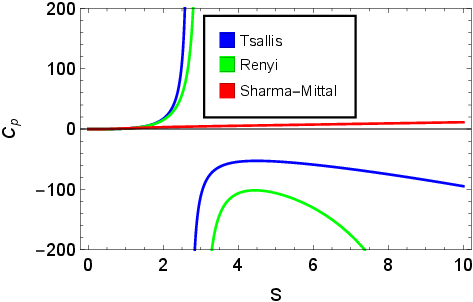,width=.45\linewidth}
\caption{$C_{p}$ versus Tsallis, R\'{e}nyi, SM entropy $S$ for
different values of cloud string $a$ (left panel) and $q$ (right panel) and  other parameters are $\beta=1$, $\delta =4$, $R = 2$, $\alpha=0.9$,  and $P = 1$, are held constant. Here, the trajectory of Tsallis, R\'{e}nyi, and  SM entropy are presented in blue, green, and red solid lines, respectively.}\label{Fig-2}
\end{figure}

Fig.~\ref{Fig-2} shows the BH-specific heat analysis using the Tsallis (blue curve), R\'{e}nyi (green curve), and SM (red curve) entropy frameworks for different values of cloud string a (left panel) and q (right panel) and other parameter include $\beta=1$, $\delta =4$, $R = 2$, $\alpha=0.9$,  and $P = 1$, are held constant. The plots show how specific heat changes with the size of the BH and its effects on stability and phase transitions. The heat capacity is an important measure of the BH thermodynamic stability. A positive heat capacity means the BH is stable and can reach thermal equilibrium with its surroundings. In contrast, a negative heat capacity means that the BH is unstable and cannot reach equilibrium. When the heat capacity is zero, it indicates a critical phase transition.
In the left panel of Fig.~\ref{Fig-2}, we examine how Tsallis (blue curve), R\'{e}nyi (green curve), and SM (red curve) entropies affect the BH thermal stability for different values of $a$. The graph shows that the SM (red curve) and R\'{e}nyi (green curve) entropy have positive heat capacities, indicating stable BH. The Tsallis (blue curve) entropy also starts with positive heat capacity, showing stability, but as $S$ increases, it undergoes a phase transition and becomes negative, indicating a change to unstable behavior. We also see that SM (red curve) is shown to be more stable than R\'{e}nyi(green entropy) entropy. In the right panel of Fig.~\ref{Fig-2}, we plot the specific heat capacity for the entropies of Tsallis (blue curve), R\'{e}nyi (green curve), and SM (red curve) at different values of $q$. Initially, the Tsallis (blue curve) and R\'{e}nyi (green curve) entropies are positive, indicating stability. After that, as $S$ increases, they undergo a phase transition at zero entropy and become negative. In contrast, the SM (red curve) entropy remains positive, reflecting stable behavior. This analysis shows that the BH behaves differently depending on which entropy framework is used. SM (red curve) entropy appears to show greater stability than both  Tsallis (blue curve) and R\'{e}nyi (green curve) entropies, especially when considering factors such as the cloud of strings $a$ and electric charge $q$. It suggests that SM entropy may be more suitable for describing BH thermodynamics under these conditions.

\subsection{The HELMHOLTZ FREE ENERGY}

Helmholtz free energy detailed in Refs.~\cite{caneva2021helmholtz, paul2024thermodynamics,simovic2024euclidean} serves as a tool to examine BHs global stability and is defined as
\begin{eqnarray}\label{436}
G=H_{M}-S \ T.
\end{eqnarray}
As we know in extended thermodynamics, BH mass can be interpreted as
the enthalpy of the system instead of internal energy. By utilizing
Eqs.~(\ref{414}), (\ref{420}) and (\ref{423}) in Eq.~(\ref{436}), then the Helmholtz free energy
of Tsallis entropy can be obtained,
\begin{eqnarray}\nonumber
F_{\mathrm{Tsallis}}&=&-\frac{1}{6 \beta}\bigg[I^{-\frac{2}{3 \beta }-1}
\left(I^{1/\beta }\right)^{\beta } \sqrt{I^{\frac{2}{3 \beta }}+q^2}
\left\{I^{\frac{2}{3 \beta }} \left(-a+8 \pi  I^{\frac{2}{3 \beta }}
P+1\right)+2 (a-1) q^2\right\}\bigg]\\\label{437}&-&\frac{1}{6}
I^{-\frac{2}{3 \beta }}\left(I^{\frac{2}{3 \beta }}+q^2\right)^{3/2}
\left(3 a-8 \pi I^{\frac{2}{3 \beta }} P-3\right).
\end{eqnarray}
Similarly, by utilizing Eqs.~(\ref{415}),~(\ref{421}) and
(\ref{424}) in Eq. (\ref{436}), then the Helmholtz free energy for R\'{e}nyi entropy  can be
derived as
\begin{eqnarray} \nonumber
F_{\mathrm{Renyi}}&=&\frac{e^{\alpha S} \sqrt{\frac{2^{2/3} N^{2/3}}{\alpha
^{2/3}}+\pi ^{4/3} q^2} \log \left(e^{\alpha  S}\right)
\bigg\{\sqrt[3]{\pi } (a-1) \alpha ^{2/3} \left(N^{2/3}-N_1\right)-8\
2^{2/3} N^{4/3} P\bigg\}}{6 \pi  \alpha ^{2/3}
N^{5/3}}\\\label{438}&+&\frac{\pi ^{4/3} \alpha ^{2/3}
\left(\frac{2^{2/3} N^{2/3}}{\pi ^{4/3} \alpha
^{2/3}}+q^2\right)^{3/2} \left\{-3 a+\frac{8\ 2^{2/3} N^{2/3}
P}{\sqrt[3]{\pi } \alpha ^{2/3}}+3\right\}}{6\ 2^{2/3} N^{2/3}}.
\end{eqnarray}
Now, by utilizing Eqs.~(\ref{416}),~(\ref{422}) and (\ref{425}) in Eq. (\ref{436}), then we obtained the Helmholtz free energy of SM entropy is given as
\begin{eqnarray}\nonumber
F_{\mathrm{SM}}&=&\frac{1}{6 \pi  R X^{5/3}}\bigg[\sqrt[3]{\delta }
\sqrt{\pi ^{4/3} q^2+\left(2 X\delta^{-1}\right)^{2/3}} \left\{(\pi
X+1)^{R/\delta }-1\right\} X_1^{\frac{\delta }{R}-1}
\bigg\{\sqrt[3]{\pi } (a-1) \delta ^{2/3}
\big(X^{2/3}\\\label{439}&-&X_2\big)-8\ 2^{2/3} P
X^{4/3}\bigg\}\bigg]+\frac{\pi ^{4/3} \delta ^{2/3}
\left(q^2+\frac{2^{2/3} X^{2/3}}{\pi ^{4/3} \delta
^{2/3}}\right)^{3/2} \left\{-3 a+\frac{8\ 2^{2/3} P
X^{2/3}}{\sqrt[3]{\pi } \delta ^{2/3}}+3\right\}}{6\ 2^{2/3}
X^{2/3}}.
\end{eqnarray}




\begin{figure} \centering
\epsfig{file=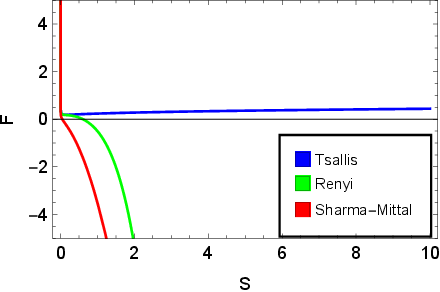,width=.45\linewidth}
\epsfig{file=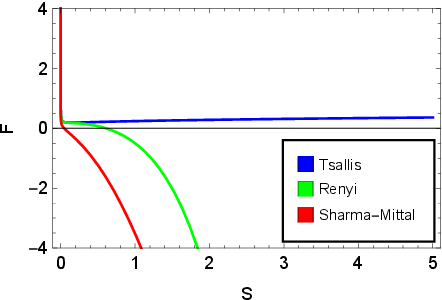,width=.45\linewidth}
\caption{The Helmholtz free energy versus Tsallis, R\'{e}nyi and  SM entropy
$S$ for different value of cloud string  $a$ (left panel) and $q$ (right panel) while other parameters are $\beta=1$, $\delta =4$, $R = 2$, $\alpha=0.9$,  and $P = 1$, are held constant. Here, the trajectory of Tsallis, R\'{e}nyi, and  SM entropy are presented in blue, green, and red solid lines, respectively.}\label{Fig-3}
\end{figure}

In Fig.~\ref{Fig-3}, we compare Helmholtz free energy for Tsallis (blue curve), R\'{e}nyi (green curve), and SM (red curve) entropies to analyze the BH stability for different values of $a$ and $q$. Positive Helmholtz free energy indicates stability, negative values suggest instability, and zero marks a phase transition. In the left panel, the Tsallis entropy (blue curve) stays positive for varying $a$, showing stability. However, the R\'{e}nyi (green curve) and SM (red curve) entropies start positive but turn negative as $S$ increases, indicating phase transitions and instability. Specifically, the R\'{e}nyi entropy becomes negative at $S=0.8$, and the SM entropy transitions at $S=0.1$. In the right panel, for different $q$, the Tsallis entropy(blue curve) again remains positive, reflecting stable behavior. Meanwhile, both the R\'{e}nyi(green curve) and SM(red curve) entropies begin positive, but R\'{e}nyi entropy and SM entropies turn negative at $S=0.9$, $S=0.01$ respectively, and signify instability for the BH as entropy increases. This analysis shows that the BH behaves differently depending on which entropy framework is used. The graph shows that Tsallis entropy shows stability, while R\'{e}nyi and SM entropies show instability for different values of $a$ and $q$. These results suggest that these entropy frameworks are useful for describing BH thermodynamics, particularly in the context of the cloud of strings and electric charge. However, R\'{e}nyi and SM entropies behave differently, showing instability under certain conditions, like the cloud of strings $a$ and charge $q$. This suggests that R\'{e}nyi and SM entropies may have limitations in describing BH systems in these specific physical contexts.

\subsection{The GIBBS FREE ENERGY}

The Gibbs free energy plays an important role in studying the BH global stability and phase transition. Detailed of the Gibbs free energy in Refs.~\cite{ali2019thermodynamics, kubizvnak2012p,deng2018thermodynamics} and is defined as
\begin{eqnarray}\label{440}
G=H_{M}-S \ T + P V.
\end{eqnarray}
To find the Gibbs free energy, we use Eqs.~(\ref{414}),~(\ref{420}), (\ref{423}), (\ref{426}) and (\ref{429}) in Eq (\ref{440}), then the Gibbs free energy of Tsallis entropy is given as,
 \begin{eqnarray}\nonumber
G_{\mathrm{Tsallis}}& =& \frac{1}{6\beta} \pi^{-2\beta} I^{-1 - \frac{2}{3\beta}} 
\sqrt{q^2 + I^{\frac{2}{3\beta}}} ( - \left( (-1 + a) q^2 \left( 2\pi^{2\beta} I + 3 \times 2^\beta S \beta \right) \right) 
- 8 P \pi^{\frac{4}{3\beta}} \left( \pi^{2\beta} \left( I - 2^{1-\beta} S \beta \right) 
+ I^{\frac{2}{3\beta}} \right) \\\label{441} &\times& \left( (-1 + a) \pi^{2\beta} \left( I - 2^\beta \times (-3 + 3a - 16P \pi q^2) S \beta \right) \right) )
\end{eqnarray}
Similarly, we use Eqs.~(\ref{415}),~(\ref{421}), (\ref{424}), (\ref{427}) and (\ref{430}) in Eq (\ref{440}), then the Gibbs free energy of R\'{e}nyi  entropy is given as,
\begin{eqnarray}\nonumber
G_{\mathrm{Renyi}} & =& \frac{1}{12 N^{5/3} \pi \alpha^{2/3}} 
\sqrt{\pi^{4/3} q^2 + \frac{2^{2/3} N^{2/3}}{\alpha^{2/3}}} ( \left( N 32 N^{1/3} P \left( -2^{2/3} + 2^{2/3} a S^\alpha 
+ N^{1/3} \pi^{4/3} q^2 \alpha^{2/3} \right) \right) 
- 3x (-1 + a) \pi^{1/3} \\\label{442}&\times& \left( 2 N^{2/3} + N_{1} \right) \alpha^{2/3} + 2 S^\alpha \left( -8 \times 2^{2/3} N^{4/3} P 
+ (-1 + a) x^{2/3} \left( N^{2/3} - N_{1} \right) \alpha^{2/3} \right) 
\log(S^\alpha))
 \end{eqnarray}
Now again similarly, by using Eqs.~(\ref{416}),~(\ref{422}), (\ref{425}), (\ref{428}) and (\ref{431}) in Eq (\ref{440}), then the Gibbs free energy of SM  entropy is given as,
\begin{eqnarray}\nonumber
 G_{\mathrm{SM}} &=& \frac{1}{12\pi R X^{5/3} \delta^{2/3}} \sqrt{\pi^{4/3} q^2 + \frac{2^{2/3} X^{2/3}}{\delta^{2/3}}} ( 16 P R X^{5/3} ( 2^{2/3} X^{2/3} \delta^{2/3} ) 
+ 2^{1/3} R X ( 8 \cdot 2^{2/3} P X^{2/3} 
- 3x (-1+a) \pi^{1/3} \\\nonumber &\times& \delta^{2/3} ) ( 2^{2/3} X^{2/3} + \pi^{4/3} q^2 \delta^{2/3} )
- 2 X_{1}^{1 - \frac{6}{R}} ( -1 + (1+\pi X)^{R/6} ) ( 8 \times 2^{\frac{2}{3}} P X^{\frac{1}{3}} + (-1 + a) \pi^{\frac{1}{3}} ( -X^{\frac{2}{3}} + 2^{\frac{1}{3}} \pi^{\frac{4}{3}} q^2 \delta^{\frac{2}{3}} \\\label{443}&\times&) \delta^{\frac{2}{3}} \delta)
\end{eqnarray}

\begin{figure} \centering
\epsfig{file=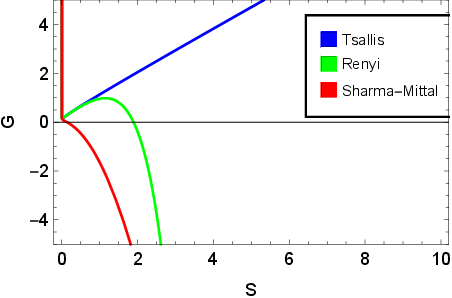,width=.45\linewidth}
\epsfig{file=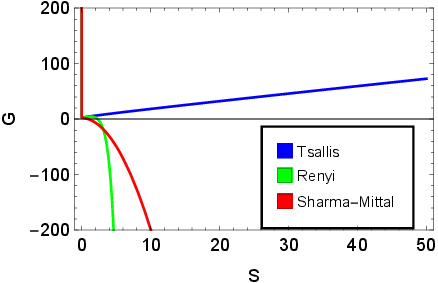,width=.45\linewidth}
\caption{The Gibbs free energy versus Tsallis, R\'{e}nyi and  SM entropy
$S$ for different value of cloud string  $a$ (left panel) and $q$ (right panel) while other parameters are $\beta=1$, $\delta =4$, $R = 2$, $\alpha=0.9$,  and $P = 1$, are held constant. Here, the trajectory of Tsallis, R\'{e}nyi, and  SM entropy are presented in blue, green, and red solid lines, respectively.}\label{Fig-4}
\end{figure}

In Figure~\ref{Fig-4}, we compare the Gibbs free energy for Tsallis (blue curve), R\'{e}nyi (green curve), and SM (red curve) entropies to study  BH stability for different values of $a$ and $q$. A positive Gibbs free energy indicates stability, a negative value suggests instability, and zero marks a phase transition. In the left panel, the Tsallis entropy (blue curve) remains positive for varying a, indicating stability. On the other hand, the R\'{e}nyi entropy (green curve) starts positive but turns negative as entropy S increases, while the SM entropy (red curve) begins at zero and becomes negative, signaling phase transitions and instability. Specifically, the R\'{e}nyi entropy becomes negative at $S=2.6$ and the SM entropy transitions at $S=0.3$. In the right panel, for different values of $q$, the Tsallis entropy (blue curve) again stays positive, reflecting stable behavior. In contrast, both the R\'{e}nyi (green curve) and SM (red curve) entropies start at zero, with the R\'{e}nyi entropy turning negative at $S=2.5$ and the SM entropy at $S=0.1$, indicating instability as entropy increases. This analysis reveals that the BH behavior depends on the entropy framework used. The Tsallis entropy consistently shows stability, while the R\'{e}nyi and SM entropies exhibit instability under certain conditions for different values of $a$ and $q$. These findings suggest that these entropy frameworks are useful for describing BH thermodynamics, the R\'{e}nyi and SM entropies may have limitations in describing BH systems under specific physical conditions, such as those involving the cloud of strings parameter $a$ and $q$.

\section{Extensive exploration of Thermal Geometries}

Geometric principles, reflected in thermodynamic geometry, have greatly enhanced our comprehension of BH thermodynamic structure. The curvature scalar is an invariant defined within this parameter space, providing deeper insights into phase transitions and the microscopic nature of BH. These concepts are proposed within the framework of thermal fluctuation theory, which has given rise to thermodynamic geometry. Usually, geothermodynamics is employed to study the interaction nature of the microstructures of the BHs.
 Now, we will examine the thermodynamic geometry associated
with the Bardeen BH. To facilitate this investigation, we employ the
geothermodynamics framework that enables the analysis of intricate
connections between thermodynamic variables and the underlying
geometric framework. Various methodologies will be explored in this analysis, such as the Weinhold, Ruppeiner, Quevedo-I, and Quevedo-II formalisms, each metric formalism possessing its advantages and limitations. Specifically, negative scalar curvatures indicate
dominant attractive micro-interactions, while positive curvatures
suggest repulsive interactions. A flat curvature signifies
non-interacting systems, like an ideal gas, or systems where
Interactions are perfectly balanced. Therefore, by selectively utilizing these diverse approaches, we aim to comprehensively address the geometric intricacies inherent in the Bardeen BH thermodynamics.
\subsection{Basic Formalisms of Geothermodynamics}
To begin with, we delve into the Weinhold geometry, a method that offers a window to represent the thermal landscape visually. We have
thoroughly examined the line element, determining the geometry as
efficiently expressed in Ref.~\cite{Jawad:2023jkb}, and it is given as
\begin{equation}\label{444}
g^{W}_{ik}=\partial_{i}\partial_{k}M(S,q),
\end{equation}
and
\begin{equation}\label{445}
ds^{2}_{W}=M_{SS}dS^{2}+M_{qq}dq^{2}+2 M_{Sq}dSdq.
\end{equation}
The two-dimensional metric for a Bardeen BH can be represented by
the following matrix
\begin{equation}\label{446}
g^{W} =
\begin{bmatrix}
M_{SS} & M_{Sq} \\
M_{qS} & M_{qq}
\end{bmatrix}.
\end{equation}
In our case, the Bardeen BH spacetime is $2$-dimensional, therefore the metric tensor elements take the given shape
$g_{SS},~g_{Sq},~g_{qS}$ and $g_{qq}$ by using the mass function.
Thereby, the components of Weinhold metrics are given below in the form of the mass function as
\begin{equation}\label{447}
M_{SS}=\frac{\partial^{2}M(S,q)}{\partial S^{2}},
M_{Sq}=\frac{\partial^{2}M(S,q)}{\partial S \partial q},
\end{equation}
\begin{equation}\label{448}
M_{qS}=\frac{\partial^{2}M(S,q)}{\partial q \partial S},
M_{qq}=\frac{\partial^{2}M(S,q)}{\partial q^{2}}.
\end{equation}
Now, we shift our focus to alternative geothermodynamic approaches
that have yielded more effective physical outcomes. In the
subsequent phase, we will examine the thermodynamic geometry of
Bardeen BH is enveloped by string clouds that employ the Quevedo
(I-II) metrics. The mathematical representation of the Quevedo
metric is as follows \cite{Jawad:2023jkb}
\begin{equation}\label{449}
 g=\bigg(E^{c} \frac{\partial \Phi}{\partial E^{c}}\bigg)\bigg(\eta_{ab}\delta^{bc}\frac{\partial^{2}\Phi}{\partial E^{c}\partial E^{d}}d E^{c}d
 E^{d}\bigg),
\end{equation}
\begin{equation}\label{450}
\frac{\partial \Phi}{\partial E^{c}}=\delta_{cb}l^{b}.
\end{equation}
The thermodynamic potential has allowed us to incorporate a wider variety of variables in the thermodynamic framework, including extensive variables such as $E^{c}$ and intensive variables such as $I^{b}$. The Ruppeiner and the Weinhold's geometric methods do not exhibit
Legendre invariance. It indicates that the Ruppeiner, and the
Weinhold metrics might yield conflicting results on occasion. To
address these inconsistencies, the Quevedo (I-II) introduced a
novel Legendre-invariant framework, which ensures that its
characteristics remain unchanged under Legendre transformations.
Within this framework, we encounter two distinct Legendre-invariant
thermogeometric metrics, the Quevedo (I-II). The groundwork for
these metrics was laid by Hermann and Mrugala
\cite{mrugala1978geometrical}, a foundation that the Quevedo (I-II) have further developed and applied. In the examination of
thermodynamic systems, the utilization of this Legendre-invariant
technique ensures consistency and compatibility.

\begin{equation}\label{451}
ds^2 =
\begin{cases}
\big(SM_S + QM_q\big)\big(-M_{SS}dS^2 + M_{qq}dq^2\big) & \text{Quevedo Case I} \\
\big(-M_{SS}dS^2 + M_{q}dq^2\big)\ SM_S & \text{Quevedo Case II}.
\end{cases}
\end{equation}
The formulation of the denominator for the curvature scalar in these
metrics is constructed as outlined in
Ref.~\cite{Soroushfar:2019ihn}
\begin{equation}\label{452}
\begin{aligned}
\text{denom } R(\text{Quevedo I}) &= 2M_{SS}^2M_{qq}^2 \big(SM_S + qM_q \big)^3, \\
\text{denom } R(\text{Quevedo II}) &= 2S^3M_{SS}^2M_{qq}^2M_S^3.
\end{aligned}
\end{equation}
A briefed investigation of the Eqs.~(\ref{441}) and (\ref{442})
demonstrates that the curvature scalar computed from the Quevedo
formalism fails to offer significant physical information regarding the system.
\subsection{Tasallis Correction}
This subsection incorporates the Tasallis entropy correction in the geothermodynamic framework to discuss the microscopic interaction between the particles of the Bardeen BH surrounded by clusters of strings. Thereby, we computed the second-order derivatives of BH's mass, which yield
\begin{eqnarray}\nonumber
M_{SS}&=&\frac{1}{18 S^{2}\sqrt{q^{2}+I^{\frac{0.66}{\beta}
}\beta^{2}}} I^{\frac{-0.66}{\beta}} \bigg[-24 P \pi
I^{\frac{2}{\beta}} (-1+ \beta) - 2 (-1+a) q^{4} (2+3
\beta)\\\label{453}&-& (-1+a) q^{2} I^{\frac{0.66}{\beta}} (2+3
\beta) + I^{\frac{1.33}{\beta}} \big\{1+8 P \pi q^{2} \big(2 -3\beta
\big)-3\beta+a(-1+3 \beta)\big\}\bigg],\\
\label{454}M_{Sq}&=&\frac{1}{ 6 S\sqrt{q^{2}+I^{\frac{0.66}{\beta}
}\beta}} I^{\frac{-0.66}{\beta}} \bigg[6(-1+a) q^{3}+q
I^{\frac{0.66}{\beta}}
\bigg\{-3+3a +8 P \pi I^{\frac{0.66}{\beta} }\bigg\}\bigg], \\
\label{455}M_{qS}&=&\frac{1}{ 6 S\sqrt{q^{2}+I^{\frac{0.66}{\beta}
}\beta}} I^{\frac{-0.66}{\beta}} \bigg[6(-1+a) q^{3}+q
I^{\frac{0.66}{\beta}} \bigg\{-3+3a + 8 P \pi I^{\frac{0.66}{\beta}
}\bigg\}\bigg], \\\label{456} M_{qq}&=& \frac{1}{2 \sqrt{q^{2}+
I^{\frac{0.66}{\beta}}}} \bigg\{I^{\frac{-0.66}{\beta}  }
I^{\frac{0.66}{\beta}  }\bigg\}\bigg\{3-3a+8 P \pi
I^{\frac{0.66}{\beta}  }\bigg\}.
\end{eqnarray}


\begin{figure}
\begin{minipage}{16pc}
\includegraphics[width=16pc]{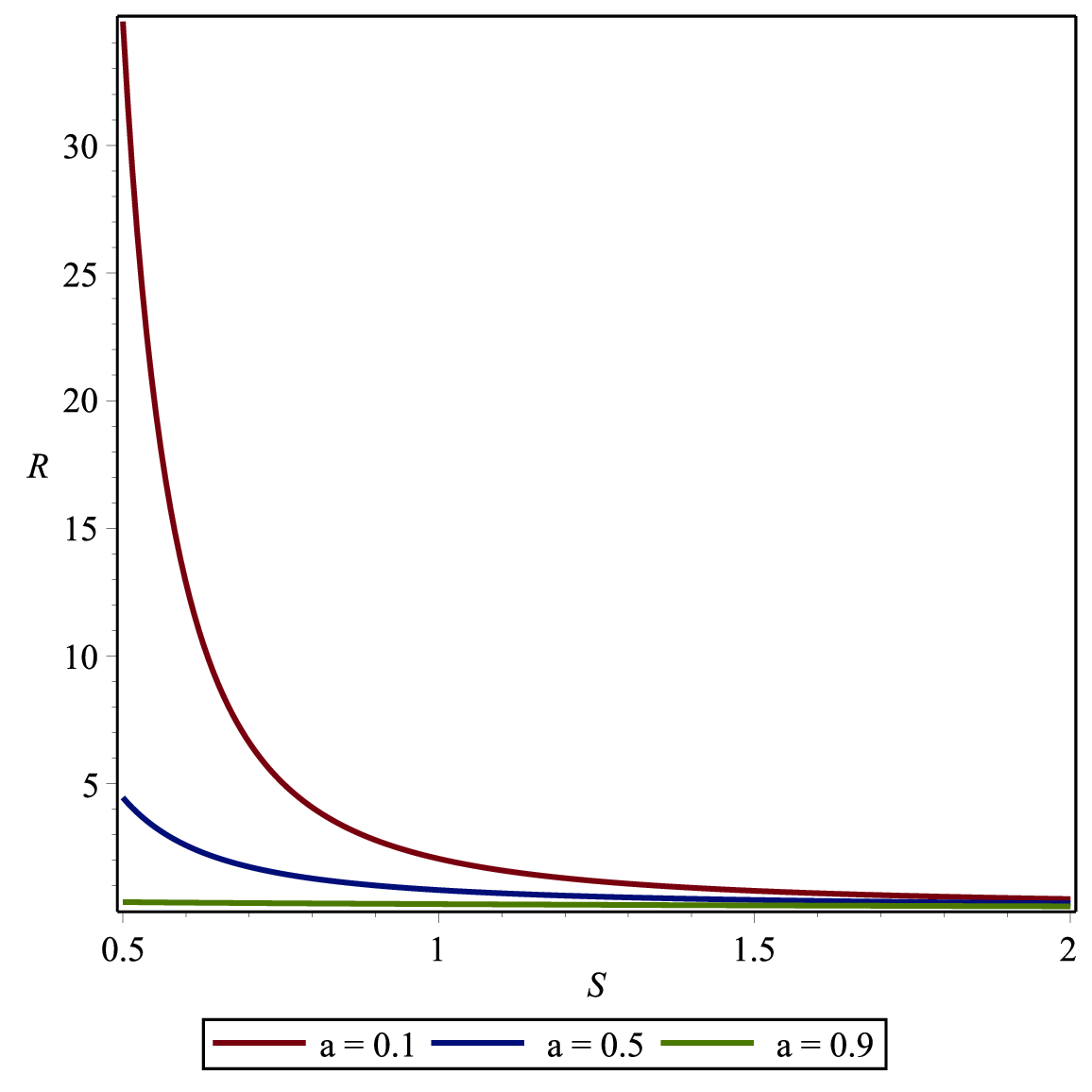}
\caption{\label{label}  Ruppeiner curvature versus Tsallis entropy
$S$ by inserting $q=0.5$, $a=0.1$ (red trajectory), $a=0.5$ (blue trajectory) and $a=0.9$ (green trajectory).}\label{Fig-13}
\end{minipage}\hspace{3pc}%
\begin{minipage}{16pc}
\includegraphics[width=16pc]{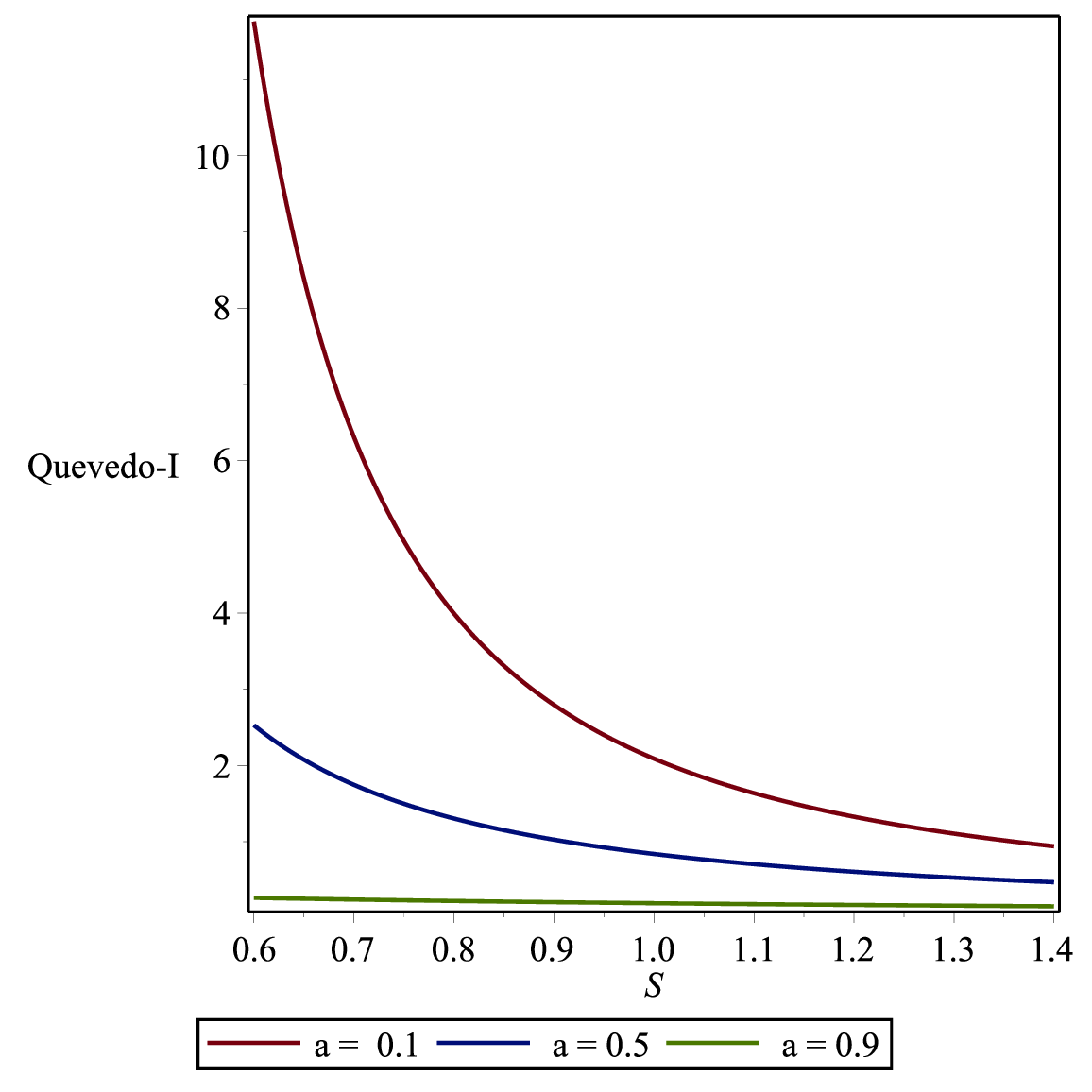}
\caption{\label{label}  Quevedo-I versus Tsallis entropy $S$ by utilizing $q=0.5$, $a=0.1$ (red trajectory), $a=0.5$ (blue trajectory) and $a=0.9$ (green trajectory).}\label{Fig-14}
\end{minipage}\hspace{3pc}%
\end{figure}
\begin{figure}
\includegraphics[width=16pc]{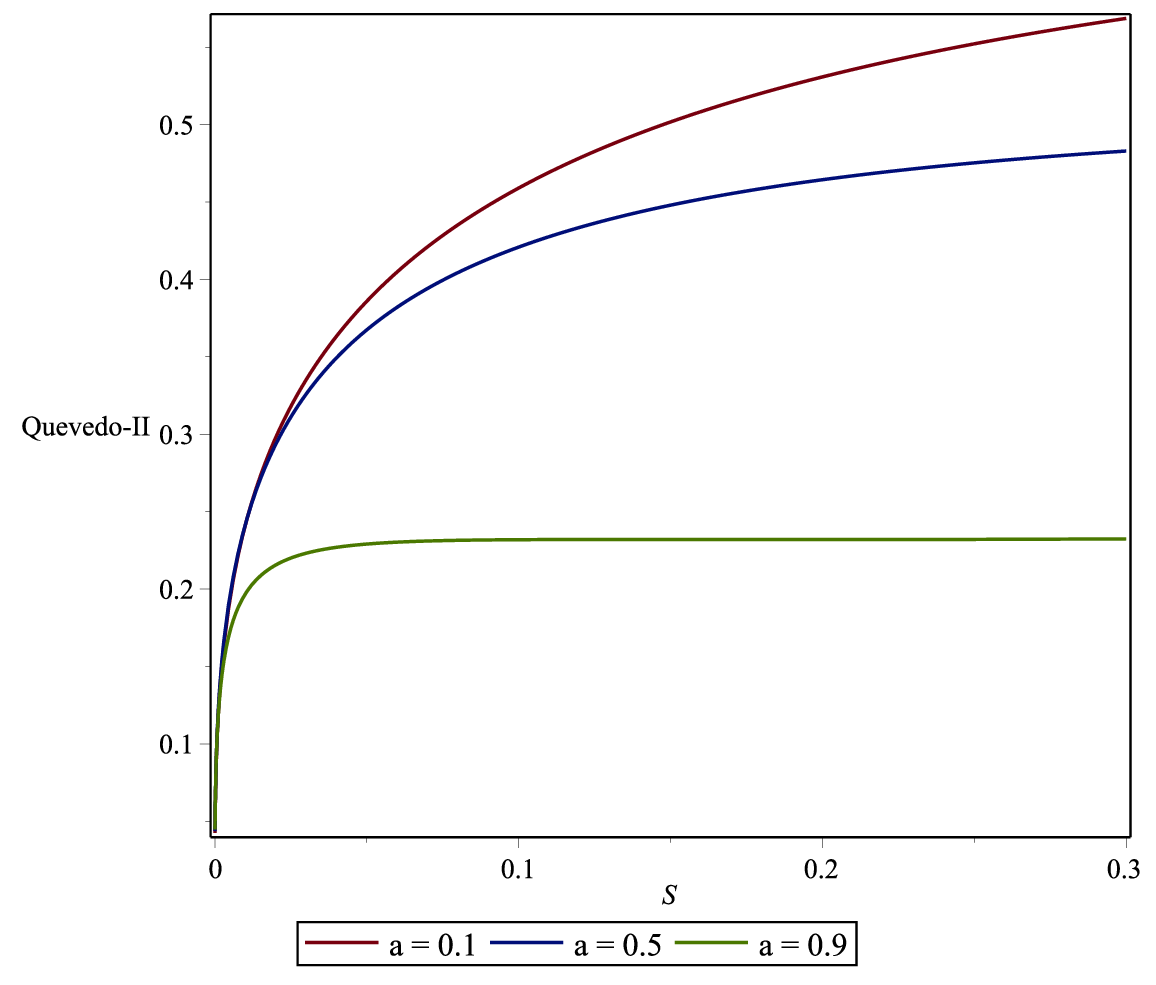}
\caption{\label{label} Quevedo-II versus Tsallis entropy $S$ by substituting  $q=0.5$, $a=0.1$ (red trajectory), $a=0.5$ (blue trajectory) and $a=0.9$ (green trajectory).}\label{Fig-15}
\end{figure}

In Fig.~\ref{Fig-13}, we plot a graph between the Ruppeiner curvature and
Tsallis entropy by inserting $q=0.5$, $a=0.1$ (red trajectory), $a=0.5$ (blue trajectory), and $a=0.9$ (green trajectory), which shows positive
behavior, indicating that microscopic structure experiences
repulsive interaction. This shows that components repel each
other, leading to a more dispersed configuration. The system shows
stability as the repulsive force prevents collapse and condensation. In
Figs.~\ref{Fig-14} and \ref{Fig-15}, we plot a graph between the Quevedo (I-II), and
Tsallis entropy by putting $q=0.5$, $a=0.1$ (red trajectory), $a=0.5$ (blue trajectory), and $a=0.9$ (green trajectory), and the graph shows
positive behavior. The positive behavior shows stability, and the
system exhibits repulsive interactions among its microstructure. The
positive Quevedo (I-II) curvature shows that the system
components tend to repel each other, leading to a more expanded,
dispersed configuration. The Quevedo-II metric formalism shows more
stable behavior than the Quevedo-I metric formalism for different values
of $a$. Let us mention here the reason why we study thermodynamic geometry formalisms for only Tsallis entropy. The
thermodynamic curvature for the Ruppeiner, Quevedo (I-II) formalisms
with Rényi and Sharma-Mittal (SM) entropies is complex, making
analytical computations impossible. However, by setting q=0, we
significantly simplify the analysis, as the thermodynamic curvature for Ruppeiner and Quevedo (I-II) in the case of Rényi and SM entropies becomes zero. This allows us to extract meaningful physical interpretations regarding BH stability and
microscopic interactions without unnecessary computational
complications. The chosen model, which is Tsallis entropy, effectively captures the influence of external parameters like a cloud of strings and charge while ensuring that the thermodynamic geometry analysis remains tractable and insightful.

 \begin{table}[ht]
\caption{\label{tab:comparison}\raggedright Summary of thermodynamic quantities by using the Tsallis, R\'{e}nyi, and SM entropies for the Bardeen BH and also its comparison with the thermodynamic quantities of the Schwarzschild BH. Here, we present the thermodynamic quantities like temperature, Heat capacity, Gibbs free energy, and Helmholtz free energy by $T,~C,~G$ and $F$, respectively.}
\begin{ruledtabular}
\begin{tabular}{cccccc}
\textbf{Quantity}              & \textbf{Tsallis} & \textbf{Rényi} & \textbf{SM} & \textbf{Bardeen BH} & \textbf{Schwarzschild BH} \\
$T$            & Stable                & Unstable               & Stable                      & Stable                & Unstable                 \\
$C$          & Unstable              & Unstable               & Stable                      & Unstable                & Unstable                 \\
$G$       & Stable                & Unstable               & Unstable                     & Stable                 & Stable                 \\
$F$  & Stable               & Unstable                & Unstable                     & Stable                & Unstable               \\
 \hline
\end{tabular}
\end{ruledtabular}
\end{table}


\section{Conclusions}

This study has presented the effects of generalized entropy on the
Bardeen BH surrounded by clusters of strings. To govern the behavior
of the string cloud, we modified the action of NED by employing the
action of Nambu-Goto. Through the examination of the Kretschmann
scalar, we explored a metric that displays singularities related to
its parameter $a$. We explored the thermal characteristics of this
solution by treating the equation of mass of the BH $M$ as the
foundational equation. We interpreted the BH mass as enthalpy by emphasizing thermodynamic pressure $P$
and entropy $S$ as crucial parameters. This approach allowed us to conduct a comprehensive
thermodynamic analysis by calculating the first-order
differentiation of enthalpy, which is associated with thermodynamic
potentials. Following standard practices in BH thermodynamics, we
evaluated the stability of the solution by examining the
thermodynamic quantities, particularly the heat capacity $C_{p}$,
under the effect of the Tsallis, R\'{e}nyi, and SM entropies. We assessed the solution stability across two different regions, which was distinguished by an unstable interval. The occurrence of an
inconsistency indicated a phase transition of first order. We
plotted a graph between the heat capacity $C_{p}$ and the Tsallis,
R\'{e}nyi, and SM entropies for different values of $a$ and $q$. We
found that graphs show stable and unstable behavior of the BH. We
provided a comprehensive analysis of the different models of
entropy. We also find the Gibbs free energy of the Tsallis, R\'{e}nyi, and SM entropies for different values of $a$. We find that the Tsallis entropy shows stable behavior, and the R\'{e}nyi
and SM entropies show unstable behavior.

Furthermore, we employed various thermodynamic geometric formalisms
to comprehend the microscopic structure of the BH. Therefore, we
investigated the Ruppeiner curvature, Quevedo (I-II), versus the Tsallis
entropy for different values of $a$, and these graphs show that
positive behavior indicates that microscopic structures experience
repulsive interaction. This shows that components repel each
other, leading to a more dispersed configuration. The positive
Quevedo (I-II) curvature shows that the system components tend to
repel each other, leading to a more expanded and dispersed
configuration. The Quevedo-II shows more stable behavior than the Quevedo-I
for different values of $a$. Furthermore, we observed that the
curvature scalar of Ruppeiner, Quevedo (I-II) in terms of R\'{e}nyi
and SM entropies became zero, which further emphasizes that by
employing Tsallis entropy, one can get better results as compared to
the R\'{e}nyi and SM entropies. Also, our results are more effective
as we have comprehensively investigated and graphically presented
the impact of various entropies. Furthermore, the Tsallis entropy might
be quite helpful as it offers more valuable insights regarding the
thermal stability of the BH, and it would be quite interesting to
observe the thermodynamic behavior of BHs through the window of
the Tsallis entropy. \vspace{2cm}

\section*{Acknowledgements}

The work of KB was supported by the JSPS KAKENHI Grant Numbers JP21K03547, 24KF0100.

\end{document}